\documentstyle[epsfig,aps,amsmath,amsfonts,amssymb,prl,multicol]{revtex}
\draft
\begin{document}
\begin{sloppypar}
\title{Testing statistical bounds on entanglement using quantum chaos}
\author {Jayendra N. Bandyopadhyay
and Arul Lakshminarayan} 
\address{Physical Research Laboratory,\\
Navrangpura, Ahmedabad 380009, India.}
\date{\today}
\maketitle
\begin{abstract}
  Previous results indicate that while chaos can lead to substantial
  entropy production, thereby maximizing dynamical entanglement, this
  still falls short of maximality. Random Matrix Theory (RMT) modeling
  of composite quantum systems, investigated recently, entails an
  universal distribution of the eigenvalues of the reduced density
  matrices. We demonstrate that these distributions are realized in
  quantized chaotic systems by using a model of two coupled and kicked
  tops. We derive an explicit statistical universal bound on
  entanglement, that is also valid for the case of unequal
  dimensionality of the Hilbert spaces involved, and show that this
  describes well the bounds observed using composite quantized chaotic
  systems such as coupled tops.
\end{abstract}
\begin{multicols}{2}
Recently, entanglement has been discussed extensively due to its
crucial role in quantum computation and quantum information theory
\cite{nielsen}. Since a quantum computer is  a many
particle system, entanglement is inevitable and even
desirable. Entanglement is important both at the hardware and software
levels of a quantum computer, as the efficiency of all proposed
quantum algorithms are based on it, hence its characterization as a
quantum resource. The many particle nature of a quantum computer
brings another phenomenon to the fore, that is {\it chaos}. Some
studies have enquired whether chaos will help or hinder in the operation
of a quantum computer \cite{dima}.  At a more basic level several
studies have explored the connections between quantum entanglement and
classical chaos \cite{furuya,sarkar1,arul1}, two phenomena that are prima 
facie uniquely quantum and classical respectively.

Such a connection between entanglement and chaos has been previously
studied with the example of an $N$-atom Jaynes-Cummings model
\cite{furuya}. It was found that the entanglement {\it rate} is
considerably enhanced if the initial wave packet was placed in a
chaotic region. In another work of similar kind, the authors have
related such rates to classical Lyapunov exponents with the help of a coupled 
kicked top model \cite{sarkar1}. Recently, one of us studied entanglement in
coupled standard maps \cite{arul1} and found that entanglement
increased with coupling strength, but after a certain magnitude of
coupling strength corresponding to the emergence of complete classical
chaos, the entanglement saturated. The saturation value depended on the
Hilbert space dimensions and was less than its maximum possible value.
This result implies that though there exists a maximum kinematical
limit for entanglement, dynamically it is not possible to create it by
using generic Hamiltonian evolutions on unentangled states. It should be
emphasized that such bounds are {\it statistical} and are more unlikely to
be violated the larger the Hilbert space dimension. 

Recent related work \cite{karol} calculates the mean entanglement of
pure states for the case $M = N$ by using a RMT model that allows
specification of the joint probability distribution of the eigenvalues
of the reduced density matrices (RDM). Below we calculate the
entanglement from a eigenvalue distribution that is valid for large
$M$ and $N$. We show that this distribution describes well those
obtained from a coupled kicked top model.  There is also some early
work that calculates the subsystem entropy for random pure states
\cite{page}. Apart from RMT simulations, we deal with an actual
quantum mechanical system and relate these results to the presence of
classical chaos, thus our results show in what context results such as
in \cite{karol,page} can be expected to be universal.

The previous studies on entanglement, in the context of chaos, were
based on pure states of bipartite systems, where the von Neumann
entropy of the RDM is a natural measure of quantum entanglement. We
will also initially consider pure states and point out in the end that
as a simple corollary we can estimate the entanglement of formation of
any density matrix as well.  Suppose that the state space of a
bipartite quantum system is ${\cal H} = {\cal H}_1 \otimes {\cal
  H}_2$, where $\mbox{dim} {\cal H}_1= N \le \mbox{dim} {\cal H}_2 =
M$, and $\mbox{dim}{\cal H}=d= NM$. If $\rho =\sum_{i} p_i |\phi_i
\rangle \langle \phi_i| $ is an ensemble representation of an
arbitrary state in $\cal{H}$, the entanglement of formation is found
by minimizing $\sum_i p_i E(|\phi_i\rangle)$ over all possible
ensemble realizations. Here $E$ is the von Neumann entropy of the RDM
of the state $|\phi_i\rangle$ belonging to the ensemble, {\em i.e.,}
its entanglement. For pure states $|\psi \rangle$ there is only one
unique term in the ensemble representation and the entanglement of
formation is simply the von Neumann entropy of the RDM.

The two RDMs of the bipartite state $|\psi\rangle$ are $\rho_1 =
\mbox{Tr}_2 ( | \psi \rangle
\langle \psi | )$ and $\rho_2 = \mbox{Tr}_1 ( | \psi \rangle \langle \psi | )$.
The Schmidt decomposition of $|\psi\rangle$ is the optimal
representation in terms of a product basis and is given by 
\begin{equation}
|\psi\rangle = \sum_{i=1}^{N} \sqrt{\lambda_i} |\phi_i^{(1)}\rangle
|\phi_i^{(2)}\rangle,
\end{equation} where $0 < \lambda_i \le 1$ are the (nonzero) eigenvalues 
of either RDMs and the vectors are the corresponding eigenvectors.
The von Neumann entropy $S_V$ is the entanglement $E(|\psi\rangle)$ given by 
\begin{equation}
S_V = - \mbox{Tr}_l (\rho_l \ln \rho_l) = -\sum_{i=1}^{N}\lambda_i 
\ln(\lambda_i),;\;\;l=1,2. 
\label{entropy}
\end{equation}

Under an arbitrary unitary evolution $\rho$ or $|\psi \rangle$ evolves
into states with changed entanglement. Quantum chaotic evolutions
eventually create large entanglement which fluctuates around the value
$\ln(\gamma N)$. The factor $\gamma$ depends only on the ratio $M/N$
and tends to unity (maximal entanglement) as $M \rightarrow \infty$.
Such evolutions lead to {\it universal} properties of the RDMs, which
is also shared by the RDMs of stationary states. This follows from the
universal near normal distributions of the complete pure state
components in any generic bases. The distribution of the eigenvalues
of RDMs $\{\lambda_i\}$ also follows from RMT results for correlation
matrices recently used in the analysis of data from financial time
series \cite{lalouxmitra}. Many important universal spectral
fluctuation properties of quantum chaotic systems have been modeled
and explained by RMT. We extend this success to the RDMs of composite
systems and consequently a description of quantum entanglement in
strongly interacting systems.

As the Hilbert space dimension and chaos have roles in this bound for
entanglement, coupled large spins are attractive models. A coupled
kicked tops model has already been used in this context \cite{sarkar1}, we 
generalize it here to include the case of unequal spins and symmetry breaking 
terms. The Hamiltonian of the coupled top system used is :
\begin{eqnarray}
H(t) &=& \frac{\pi}{2} J_{y_1} + \frac{k}{2j_1} (J_{z_1} + \alpha_1)^2 
\sum_{n = -\infty}^{\infty} \delta (t-n)\nonumber\\ 
&+& \frac{\pi}{2} J_{y_2} 
+ \frac{k}{2j_2} (J_{z_2} + \alpha_2)^2 \sum_{n = -\infty}^{\infty} 
\delta (t-n)\nonumber\\  
&+& \frac{\epsilon}{\sqrt{{j_1}{j_2}}} J_{z_1} J_{z_2} 
\sum_{n = -\infty}^{\infty} \delta (t-n).
\end{eqnarray}
\noindent The $J_{y_r}$ terms describe free precession of each top and the 
remaining terms are due to periodic $\delta$-function kicks. The first
two such terms are torsion about $z-$axis and the final term
describes the spin-spin coupling. When either of the constants, $\alpha_1$
or $\alpha_2$, is not zero the parity symmetry $R H(t) R^{-1} = H(t)$, 
where $R = \exp(i \pi J_{y_1}) \otimes \exp(i \pi J_{y_2})$, is broken. The
dimensionality of the Hilbert spaces are $N=2j_1+1$ and $M=2j_2+1$. The unitary 
time evolution operator corresponding to this Hamiltonian is given by :
\begin{equation}
U_T = ( U_1 \otimes U_2 )\, U_{12}^{\epsilon} = [\, ( U_{1}^{f} U_{1}^{k} ) 
\otimes ( U_{2}^{f} U_{2}^{k} ) \,] \,U_{12}^{\epsilon}, 
\end{equation}
\noindent where the different terms are given by
\begin{eqnarray}
U_{i}^{f} &\equiv& \exp \left( -\frac{i\pi}{2} J_{y_i} \right), 
U_{i}^{k} \equiv \exp \left( -\frac{ik}{2j_i} (J_{z_i} + \alpha_i)^2 \right),
\nonumber\\
U_{12}^{\epsilon} &\equiv& \exp \left( -\frac{i \epsilon}{\sqrt{j_1 j_2}} 
J_{z_1} J_{z_2} \right)
\end{eqnarray}
\noindent and $i = 1, 2$. There exists an antiunitary generalized time 
reversal symmetry, $[ \exp (i \pi J_{x_1}) \exp (i \pi J_{y_1}/2) ] \otimes [ 
\exp(i \pi J_{x_2}) \exp (i \pi J_{y_2}/2) ] K$ where $K$ is complex 
conjugation operator, from which we can expect the applicability of results
concerning the Gaussian orthogonal ensemble (GOE). We note that for 
the parameter values considered in this Letter, the 
nearest neighbor spacing distribution (NNSD) of the eigenangles of $U_T$ is 
Wigner distributed, which is typical of quantized chaotic systems with time 
reversal symmetry. Entanglement production of time evolving states
under $U_T$ have been studied for two different initial states.  (1)
The initial state is a product of directed angular momentum states as
given in Ref. \cite{sarkar1}, placed in the chaotic sea of phase
space. This is a completely unentangled state. (2) The initial
state is maximally entangled and is given by:
\begin{equation}
\langle m_1,m_2|\psi(0) \rangle = \frac{1}{N} \delta_{m_1 m_2}.
\label{ent_ini}
\end{equation}
\noindent These initial states are evolved under $U_T$, and the results
are displayed in Fig. \ref{FIG1}. Here the coupling strength is
very large compared to the value taken in \cite{sarkar1}, as our
goal is to study entanglement saturation and strong coupling will
help us achieve entanglement saturation within a short time.
In the first case, initially both the von Neumann entropy and the
linearized entropy ($ S_R = 1 - \mbox{Tr}_1( \rho_{1}^{2} )$), are zero, but
with time evolution both entropies start   
{\narrowtext
\begin{figure}
\centerline{\psfig{figure=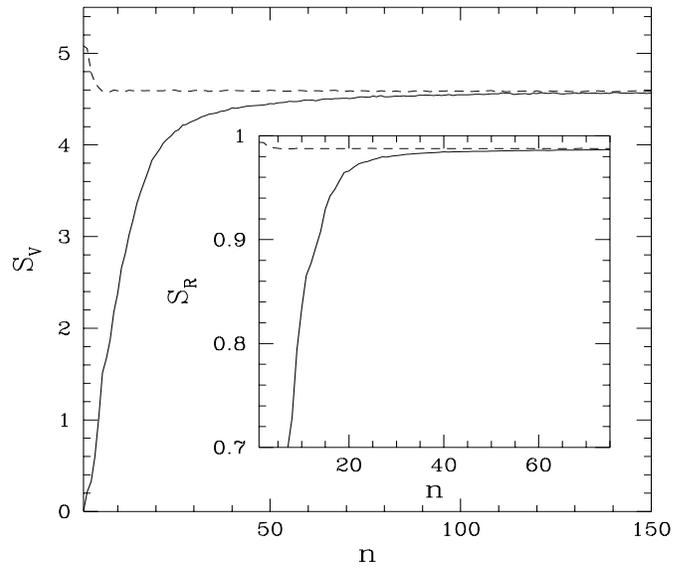,height=8cm,width=9cm}}
\caption{Entanglement saturation of a completely unentangled initial
state (solid line) and a maximally entangled initial state (dotted
line) under time evolution operator $U_T$. Here $k = 3, \epsilon = 0.1$ and
the phases $\alpha_1 = \alpha_2 = 0.47$. Inset shows similar behaviour of 
linear entropy.}
\label{FIG1}
\end{figure}}
\noindent increasing and get saturated, apart from small fluctuations, at 
values less than their maximum possible values.

For the von Neumann entropy the saturation value is $\sim \ln (0.6N)$ and for
the linear entropy it is approximately $1 - 2/N$, where $N$ is the
dimension of each subsystems. This is the dynamical bound for
entanglement of a system consists of two equal dimensional subsystems,
while the maximum kinematical limits are $\ln N$ and $1 - 1/N$
respectively. The saturation value of von Neumann entropy of this
time evolved state is same as that obtained for stationary states of
completely chaotic coupled standard maps \cite{arul1}. In the second
case, the initial state is maximally entangled and time evolution
forces this state to partially disentangle till the entropy reaches
the above mentioned values.

This study shows that the saturation of entanglement is a universal
phenomenon, it depends only on the Hilbert space dimensions, and not
on dynamical characteristics of the system, apart from the presence of
complete chaos. The effect of dimension on entanglement saturation has
been studied by keeping the dimension of the first
subsystem constant at $N$ and increasing the dimension $M$
of the second subspace from $M = N$ to some large value.
Thus we may think of the second spin as tending towards a complex
bath with a quasi-continuous spectrum. It is observed that the entanglement 
saturation increases with $M$ and finally gets saturated at the maximum 
possible  
{\narrowtext
\begin{figure}
\centerline{\psfig{figure=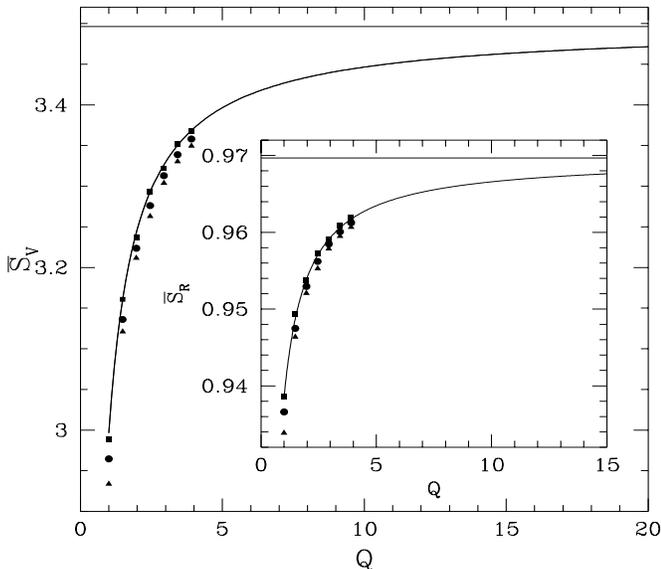,height=8cm,width=9cm}}
\caption{The spectral average of the entanglement present in eigenstates 
of $U_T$ ($k=9, \epsilon=10$) as a function of $Q = M/N$, where $N = 2j_1 + 1 
= 33$. Solid triangles are kicked top results with parity symmetry
($\alpha_1 = \alpha_2 = 0$) and solid circles are the corresponding results 
without symmetry ($\alpha_1 = \alpha_2 = 0.47$). Solid 
squares are the result of corresponding RMT Monte Carlo simulations and solid 
line is the theoretical curve Eq. (\ref{sv}). Horizontal line is the maximum 
possible entanglement ($\ln (N)$). Inset shows the behaviour of the linear 
entropy.}
\label{FIG2}
\end{figure}}
{\narrowtext
\begin{figure}
\centerline{\psfig{figure=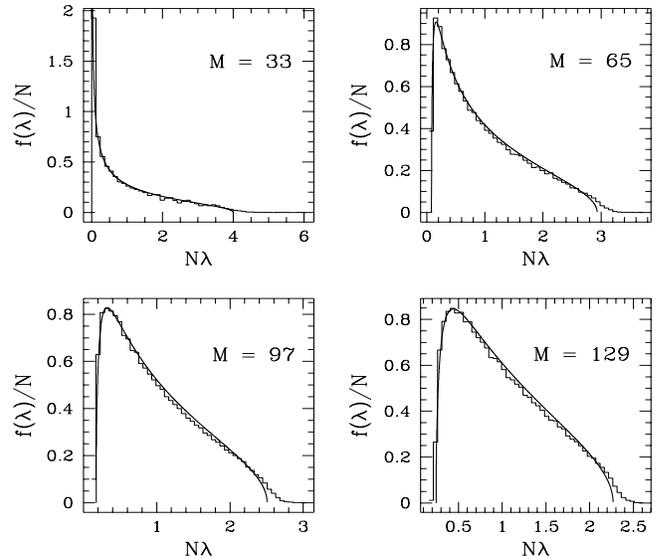,height=8cm,width=9cm}}
\caption{Distribution of the eigenvalues of the
RDMs of coupled kicked tops, averaged over all the eigenstates
($N = 2j_1 + 1 = 33$). Solid curves corresponds to the theoretical distribution 
function Eq. (\ref{flambda}).}
\label{FIG3}
\end{figure}}
\noindent kinematical limit, as shown in Fig. \ref{FIG2}. For example the von 
Neumann entropy starting from $\ln (0.6N)$ increases asymptotically to  
$\ln N$, while the linear entropy starting from $1 - 2/N$ tends to $1 - 1/N$.

We can develop a complete analytical understanding of these limits via RMT 
modeling. A typical stationary state of a quantum 
chaotic system shares properties of the eigenvectors of random matrices. 
Let us assume that some product basis has been used to write components
$a_{nm}$ of any state, which is real for stationary states of time reversal 
symmetric systems.
Writing $a_{nm}$ as the rectangular $N \times M$ matrix $A$, the $M$
dimensional RDM is $A^TA$ while the other RDM is the $N$ dimensional
$A A^T$. The assumptions of quantum chaos, we have just seen, imply
that $A$ can be taken to have random independent entries, a member of
the Laguerre ensemble. The RDMs then have the structure of correlation
matrices \cite{lalouxmitra}, from where we
directly use results for the density of states. Such matrices 
have also been studied since the early days of RMT as they have a
non-negative spectrum \cite{brody}. The distribution of the eigenvalues 
of such matrices is known and thus this is the distribution of the eigenvalues 
of RDMs. The density of the eigenvalues of the RDM $\rho_{1}$ is given by
\begin{eqnarray}
f( \lambda ) &=& \frac{NQ}{2\pi} \frac{\sqrt{( \lambda_{max} - \lambda )
( \lambda - \lambda_{min} )}}{\lambda} \nonumber\\
\lambda_{min}^{max} &=& \frac{1}{N} \left( 1 + \frac{1}{Q} \pm 
\frac{2}{\sqrt{Q}} \right),
\label{flambda}
\end{eqnarray}
\noindent where $\lambda \in[ \lambda_{min}, \lambda_{max} ]$, $Q = M/N$
and $N f(\lambda) d\lambda$ is the number of eigenvalues within $\lambda$ to
$\lambda + d\lambda$.
This has been derived under the assumption that both $M$ and $N$ are
large. Note that this predicts a range of eigenvalues for the RDMs
that are of the order of $1/N$.  For $Q \ne 1$, the eigenvalues of the
RDMs are bounded away from the origin, while for $Q=1$ there is a
divergence at the origin. All of these predictions are seen to be borne
out in numerical work with coupled tops.

Fig. \ref{FIG3} shows how well the above formula fits the eigenvalue
distribution of actual reduced density matrices. This figure also
shows that the probability of getting an eigenvalue outside the range
$[ \lambda_{min}, \lambda_{max} ]$ is indeed very small.  The sum in
$S_V$ can be replaced by an integral over the density $f(\lambda)$:
\begin{equation}
S_V \sim - \int_{\lambda_{min}}^{\lambda_{max}} f(\lambda) 
\lambda \ln \lambda \, d \lambda \,\equiv\, \ln (\gamma N)
\label{sv}
\end{equation}
\noindent The integral in $\gamma$ can be evaluated to
a generalized hypergeometric function and the final result is :
\begin{eqnarray}
\gamma &=& \frac{Q}{Q+1} \exp\left[\frac{Q}{2(Q+1)^2} ~{{_3}F_{2}} \left\{
1, 1, \frac{3}{2}; 2, 3; \frac{4Q}{(Q+1)^2} \right\} \right]
\nonumber\\
\end{eqnarray}
\noindent When the two subsystems are of equal dimension, that is $Q = 1$, then
above expression gives $\gamma = \exp(-0.5) \sim 0.6$ and so the
corresponding von Neumann entropy is $\ln (0.6N)$. This is also the
saturation value obtained in previous numerical
work for the stationary states and time evolving states of a coupled standard 
map \cite{arul1}, reflecting universality. In another
extreme case, when the Hilbert space dimension of the second subsystem is
very large compared to that of the first, that is $Q \gg 1$, 
then $\gamma \sim 1$ and hence the corresponding von Neumann entropy
is $\ln (N)$. Therefore, the analytical formulation based on RMT is able to
explain the saturation behaviour of the von Neumann entropy or quantum
entanglement very accurately. 

Fig. \ref{FIG2} also compares the
Eq. (\ref{sv}) to both RMT simulations and kicked top results. We
expect that the deviations of the quantum calculations are due to 
finite size effects. The presence of parity
symmetry results in a somewhat smaller entanglement, as seen in this figure, a
fact that needs further study. For time evolving states and stationary states
of system without time reversal symmetry the RDMs are complex Hermitian
matrices. The entanglement bounds discussed here are also valid for these cases
as the entanglement depends only on the density of states of the RDMs. 
However, spectral fluctuations of the RDMs (such as {\em their} NNSD) 
corresponding to these states can be distinct. Indeed, in the correlation 
matrix approach to atmospheric data, such a difference has been recently noted
\cite{santha}.      

The linear entropy can also be derived as above, but using direct
RMT results, without taking recourse to the distribution above,
is also possible in this case. Thus we may write:
\begin{eqnarray}
\mbox{Tr}\rho^2 
&=& \sum_{j,k=1}^{N} \sum_{\alpha,\beta=1}^{M} 
a_{j\alpha}a_{k\alpha} a_{k\beta}a_{j\beta} 
\label{tr_rhosq}
\end{eqnarray}
\noindent Substituting RMT ensemble average values \cite{ullah} of 
$\langle a_{j\alpha}^2 a_{j\beta}^{2}\rangle = \langle a_{j\alpha}^2 
a_{k\alpha}^{2}\rangle = 1/[ MN (MN+2) ]$, 
$\langle a_{j\alpha}^{4}\rangle = 3/[ MN (MN+2) ]$ and 
$\langle a_{j\alpha} a_{k\alpha}a_{k\beta}a_{j\beta}\rangle = 0$, where 
$j \neq k$ and $\alpha \neq \beta$ in the above expressions, we find that: 
\begin{equation}
S_R = 1 - \mbox{Tr} \rho_{1}^2 = 1 - \frac{M + N + 1}{MN + 2}.
\end{equation} 
\noindent When the dimension of the two subsystems are equal, that is $M = N$, 
then in the large $N$ limit $S_R \sim 1 - 2/N$. This is the saturation value of
the linear entropy approximately obtained in case of time evolving
states of coupled kicked tops. Similarly, when the Hilbert space dimension
of the second subsystem is very large compared to the dimension of first
subsystem, that is $M \gg N$,  $S_R \sim 1 - 1/N$. This is the
maximum possible value of linear entropy. 

Finally as an almost trivial corollary we note that the entanglement
of formation of a time evolving density matrix is also statistically
bounded to $\ln(\gamma N)$, as each pure state belonging to an
ensemble representation evolves to this entanglement under quantum
chaos.  To summarize, we have pointed out that the eigenvalue
distribution of reduced density matrices of composite quantum chaotic
bipartite systems are universal, and shown that there exists a typical
value of quantum entanglement that quantum chaos engenders. This value
is the maximum we may expect typical unentangled initial states to be
able to reach under generic interactions. If we already had maximally
entangled states, then chaos can disentangle this state to just such
an extent as to coincide with this generic value.

\thanks{We wish to thank Dr. M. S. Santhanam, Prof. V. K. B. Kota and
Prof. V. B. Sheorey for useful discussions. We thank Prof. K. Zyczkowski 
for informing us of references \cite{karol} and \cite{page}.}

\end{multicols}

\end{sloppypar}

\end{document}